\documentclass[conference]{IEEEtran}
\IEEEoverridecommandlockouts
\usepackage{cite}
\usepackage{amsmath,amssymb,amsfonts}

\usepackage{hyperref}
\usepackage{graphicx}
\usepackage{fontawesome5}
\usepackage{tablefootnote}
\usepackage{textcomp}
\usepackage{xcolor}
\def\BibTeX{{\rm B\kern-.05em{\sc i\kern-.025em b}\kern-.08em
    T\kern-.1667em\lower.7ex\hbox{E}\kern-.125emX}}

\usepackage{amsmath,amssymb,amsfonts}
\usepackage{graphicx}
\usepackage{textcomp}
\usepackage[most]{tcolorbox} 
\def\BibTeX{{\rm B\kern-.05em{\sc i\kern-.025em b}\kern-.08em
    T\kern-.1667em\lower.7ex\hbox{E}\kern-.125emX}}

\usepackage[most]{tcolorbox} 
\usepackage{xspace} 
\usepackage[normalem]{ulem}  
\useunder{\uline}{\ul}{}  
\usepackage{booktabs} 
\usepackage{multirow} 
\usepackage{subcaption} 
\usepackage[noend]{algpseudocode}
\usepackage{enumitem}
\usepackage{adjustbox}
\usepackage{wrapfig}
\usepackage[ampersand]{easylist}
\usepackage[most]{tcolorbox}
\newtheorem{definition}{Definition}
\usepackage{colortbl}

\definecolor{main}{HTML}{5989cf}    
\definecolor{sub}{HTML}{cde4ff}     

\newtcolorbox{boxB}{
    fontupper = \bf\color{main}\footnotesize, 
    boxrule = 0.5pt,
    colframe = main,
    rounded corners,
    arc = 5pt   
}

\newtcolorbox{boxD}{
    fontupper = \small, 
    colback = sub, 
    colframe = main, 
    boxrule = 0pt, 
    toprule = 2pt, 
    bottomrule = 2pt 
}

\newtcolorbox{boxH}{
    fontupper = \small, 
    colback = sub, 
    colframe = main, 
    boxrule = 0pt, 
    leftrule = 6pt 
}

\newtcolorbox{boxG}{
    enhanced,
    boxrule = 0pt,
    colback = sub,
    borderline west = {1pt}{0pt}{main}, 
    borderline west = {0.75pt}{2pt}{main}, 
    borderline east = {1pt}{0pt}{main}, 
    borderline east = {0.75pt}{2pt}{main}
}    

\newtcolorbox{boxK}{
    fontupper = \small,
    sharpish corners, 
    boxrule = 0pt,
    toprule = 1.0pt, 
    enhanced,
    fuzzy shadow = {0pt}{-2pt}{-0.5pt}{0.5pt}{black!35} 
}

\begin{document}

\definecolor{MidnightBlue}{HTML}{006895}

\newcommand{\projtodo}[1]{{\color{red}{\bf TODO:}#1}}
\newcommand{\JSgruCloneIII}{{$0.81$}\xspace}
\newcommand{\JStfCloneIII}{{$1.0$}\xspace}

\newcommand*\circled[1]{\tikz[baseline=(char.base)]{
            \node[shape=circle,draw,inner sep=0.5pt] (char) {#1};}}

\newboolean{showcomments}

\makeatletter 
\newcommand{\linebreakand}{%
  \end{@IEEEauthorhalign}
  \hfill\mbox{}\par
  \mbox{}\hfill\begin{@IEEEauthorhalign}
}
\makeatother 

\setboolean{showcomments}{true}

\ifthenelse{\boolean{showcomments}}
  {\newcommand{\nb}[2]{
    \fbox{\bfseries\sffamily\scriptsize#1}
    {\sf\small$\blacktriangleright$\textit{#2}$\blacktriangleleft$}
   }
   \newcommand{\cvsversion}{\emph{\scriptsize$-$Id: macro.tex,v 1.9 2005/12/09 22:38:33 giulio Exp $}}
  }
  {\newcommand{\nb}[2]{}
   \newcommand{\cvsversion}{}
  }

\newcommand\myparagraph[1]{\noindent\underline{\bf {#1}:}}
\newcommand\myparagraphnew[1]{\noindent{\bf {#1}:}}
\newcommand\emparagraph[1]{\noindent {\em {#1}:}}
\newcommand\budget[1]{{\color{red}\myparagraph{Budget}{#1} pages}}
\newcommand\KEVIN[1]{{\color{ForestGreen} \nb{KEVIN}{#1}}}
\newcommand\DANIEL[1]{{\color{red} \nb{DANIEL}{#1}}}
\newcommand\DAVID[1]{{\color{blue} \nb{DAVID}{#1}}}
\newcommand\DENYS[1]{{\color{blue} \nb{DENYS}{#1}}}
\newcommand\ALEJO[1]{{\color{purple} \nb{ALEJO}{#1}}}
\newcommand\TODO[1]{{\color{red} \nb{TODO}{#1}}}

\newcommand{\cancel}[1]{{\leavevmode\color{RubineRed}{\sout{\xspace#1}}}}
\newcommand{\edit}[2]{{\leavevmode\color{RubineRed}{\sout{#1}}}{\color{blue}{\xspace#2}}}
\newcommand{\rewrite}[2]{{\leavevmode\color{RubineRed}{\sout{#1}}}{\color{Green}{\arrow\xspace#2}}}

\newcommand{\add}[1]{{\leavevmode\color{blue}{\xspace#1}}}
\newcommand{\addnew}[1]{{\leavevmode\color{Green}{\xspace#1}}}
\newcommand{\remove}[1]{{\leavevmode\color{red}{\xspace#1}}}

\newcommand\finding[1]{\vspace{0.25em}\noindent\textsf{\bf Finding {#1}.}}
\newcommand\fnumber[1]{{$\mathcal{F}_{#1}$}}
\newcommand\operator[2]{{\bf OP$_{#1}$: {\em {#2}} -- }}
\newcommand\opnumber[1]{{{\bf OP}$_{#1}$}}

\newcommand{\arrow}{{$\rightarrow$}\xspace}
\newcommand\inline[1]{{\lstinline{#1}}}


\newcommand{\boxme}[1]{{
\begin{tcolorbox}[enhanced,skin=enhancedmiddle,borderline={1mm}{0mm}{MidnightBlue}]
    \textbf{Take Aways: } #1 \end{tcolorbox} 
}}

\newcommand\fix[1]{{\color{blue} \nb{FIX THIS}{#1}}}
\newcommand\blue[1]{{\color{blue}{#1}}}
\newcommand{\here}{{\color{blue} \nb{***}{CONTINUE HERE}}}

\newcommand{\REF}{{\color{red} \textbf{[REFS]}}\xspace}
\newcommand{\xy}{{\color{red} \textbf{XY}}\xspace}
\newcommand\tops[1]{{\color{blue}{#1}}}
\newcommand\alert[1]{{\color{red}{#1}}}

\newcommand{\target}{\textit{target tool}\xspace}
\newcommand{\targets}{\textit{target tools}\xspace}
\newcommand{\behavior}{\textit{target behavior}\xspace}
\newcommand{\ie}{\textit{i.e.,}\xspace}
\newcommand{\eg}{\textit{e.g.,}\xspace}
\newcommand{\etc}{\textit{etc.}\xspace}
\newcommand{\etal}{et al.\xspace}
\newcommand{\etals}{et al.'s\xspace}
\newcommand{\aka}{\textit{a.k.a.}\xspace}	

\newcommand{\astxplainer}{AST\textit{xplainer}\xspace}

\newcommand{\asceval}{AsC\textit{-Eval}\xspace}
\newcommand{\asccausal}{AsC\textit{-Causal}\xspace}
\newcommand{\ascviz}{AsC\textit{-Viz}\xspace}

\newcommand{\asc}{\textit{AsC}\xspace}
\newcommand{\ascs}{\textit{AsCs}\xspace}

\newcommand{\ntp}{\textit{NtP}\xspace}

\newcommand{\codegen}{\textit{ScP}\xspace}
\newcommand{\ct}{\textit{c\&t}\xspace}

\newcommand{\nlms}{NCMs\xspace}
\newcommand{\nlm}{NCM\xspace}

\newcommand{\llms}{LLMs\xspace}
\newcommand{\llm}{LLM\xspace}

\newcommand{\datainterI}{\textit{ProgramRepair}\xspace}
\newcommand{\datainterII}{\textit{SemanticPreserving}\xspace}
\newcommand{\datainterIII}{\textit{UnCommenting}\xspace}

\newcommand{\modelinterI}{\textit{NumberLayers}\xspace}
\newcommand{\modelinterII}{\textit{NumberUnits}\xspace}

\newcommand{\assoJS}{JS Dist.\xspace}
\newcommand{\assoPR}{Pearson\xspace}


\newcommand{\rfi}{$\mathcal{R}_1$\xspace}

\newcommand{\galeras}{\textit{Galeras}\xspace}
\newcommand{\dataset}{\textit{CodeSearchNet}\xspace}

\newcommand{\blocks}{\texttt{\small[blocks]}\xspace}
\newcommand{\tests}{\texttt{\small[tests]}\xspace}
\newcommand{\oop}{\texttt{\small[oop]}\xspace}
\newcommand{\declarations}{\texttt{\small[declarations]}\xspace}
\newcommand{\exceptions}{\texttt{\small[exceptions]}\xspace}
\newcommand{\datatype}{\texttt{\small[datatype]}\xspace}
\newcommand{\loops}{\texttt{\small[loops]}\xspace}
\newcommand{\operators}{\texttt{\small[operators]}\xspace}
\newcommand{\conditionals}{\texttt{\small[conditionals]}\xspace}
\newcommand{\extra}{\texttt{\small[extraTokens]}\xspace}

\newcommand{\gptI}{\textit{gpt-3 [125M]}\xspace}
\newcommand{\gptII}{\textit{gpt-3 [1.3B]}\xspace}
\newcommand{\gptIII}{\textit{gpt-3 [2.7B]}\xspace}

\newcommand{\codegenI}{\textit{codegen-nl [350M]}\xspace}
\newcommand{\codegenII}{\textit{codegen-nl [2B]}\xspace}

\newcommand{\multiI}{\textit{multi-lang [110M]}\xspace}
\newcommand{\multiII}{\textit{multi-lang [350M]}\xspace}
\newcommand{\multiIII}{\textit{multi-lang [2B]}\xspace}

\newcommand{\monoI}{\textit{mono-lang [110M]}\xspace}
\newcommand{\monoII}{\textit{mono-lang [1.5B]}\xspace}
\newcommand{\monoIII}{\textit{mono-lang [350M]}\xspace}
\newcommand{\monoIIII}{\textit{mono-lang [2B]}\xspace}

\newcommand{\nlgpt}{\textit{NL GPT-3}\xspace}
\newcommand{\nlcodegen}{\textit{NL Codegen}\xspace}
\newcommand{\monolang}{\textit{Mono-Language-Type}\xspace}
\newcommand{\multilang}{\textit{Multi-Language-Type}\xspace}
\newcommand{\COVwmd}{\texttt{\small(wmd)}\xspace} 
\newcommand{\COVloc}{\texttt{\small(LoC)}\xspace} 
\newcommand{\COVreturn}{\texttt{\small(\#returns)}\xspace} 
\newcommand{\COVloop}{\texttt{\small(\#loops)}\xspace} 
\newcommand{\COVcomparison}{\texttt{\small(\#comparisons)}\xspace} 
\newcommand{\COVtry}{\texttt{\small(\#tryCatches)}\xspace} 
\newcommand{\COVparenthesized}{\texttt{\small(\#parenthesized)}\xspace} 
\newcommand{\COVexpression}{\texttt{\small(\#expressions)}\xspace} 
\newcommand{\COVnumber}{\texttt{\small(\#numbers)}\xspace} 
\newcommand{\COVstring}{\texttt{\small(\#stringLiterals)}\xspace} 
\newcommand{\COVmathops}{\texttt{\small(\#mathOps)}\xspace} 
\newcommand{\COVvari}{\texttt{\small(\#variables)}\xspace} 
\newcommand{\COVmnestedblock}{\texttt{\small(\#maxNextedBlocks)}\xspace} 
\newcommand{\COVanonyclass}{\texttt{\small(\#anonyClasses)}\xspace} 
\newcommand{\COVinnerclass}{\texttt{\small(\#innerClasses)}\xspace} 
\newcommand{\COVlambdaexp}{\texttt{\small(\#lambdaExpressions)}\xspace} 

\newcommand{\COVunique}{\texttt{\small(\#uniqueWords)}\xspace} 
\newcommand{\COVlog}{\texttt{\small(\#logStatements)}\xspace} 
\newcommand{\COVmod}{\texttt{\small(\#modifiers)}\xspace}

\newcommand{\secref}[1]{Sec.~\ref{#1}\xspace}
\newcommand{\figref}[1]{Fig.~\ref{#1}\xspace}
\newcommand{\tabref}[1]{Table~\ref{#1}\xspace}
\newcommand{\Phase}{{\sc Phase}\xspace}
\newcommand{\Phases}{{\sc Phase's~}\xspace}

\newcommand{\emphquote}[1]{{\emph{`#1'}}\xspace}
\newcommand{\emphdblquote}[1]{{\emph{``#1''}}\xspace}

\newcommand{\emphbrack}[1]{\emph{[#1]}\xspace}
			
\newcommand{\subj}{\emphbrack{subject}}
\newcommand{\act}{\emphbrack{action}}
\newcommand{\obj}{\emphbrack{object}}
\newcommand{\prep}{\emphbrack{preposition}}
\newcommand{\objtwo}{\emphbrack{object2}}

\newcommand*\ciclednum[1]{\raisebox{.5pt}{\textcircled{\raisebox{-.9pt}
{#1}}}}

\bstctlcite{IEEEexample:BSTcontrol}

\title{
Evaluating and Explaining Large Language Models for Code Using Syntactic Structures
}


\author{\IEEEauthorblockN{1\textsuperscript{st} David N.~Palacio}
\IEEEauthorblockA{\textit{Department
of Computer Science} \\
\textit{William \& Mary}\\
Williamsburg, VA \\
danaderpalacio@wm.edu}
\and
\IEEEauthorblockN{2\textsuperscript{nd} Alejandro Velasco}
\IEEEauthorblockA{\textit{Department
of Computer Science} \\
\textit{William \& Mary}\\
Williamsburg, VA \\
svelascodimate@wm.edu}
\and
\IEEEauthorblockN{3\textsuperscript{rd} Daniel Rodriguez-Cardenas}
\IEEEauthorblockA{\textit{Department
of Computer Science} \\
\textit{William \& Mary}\\
Williamsburg, VA \\
dhrodriguezcar@wm.edu}
\and

\linebreakand 

\IEEEauthorblockN{4\textsuperscript{th} Kevin Moran }
\IEEEauthorblockA{\textit{Department of Computer Science} \\
\textit{University of Central Florida}\\
Orlando, FL \\
kpmoran@ucf.edu}
\and
\IEEEauthorblockN{5\textsuperscript{th} Denys Poshyvanyk}
\IEEEauthorblockA{\textit{Department
of Computer Science} \\
\textit{William \& Mary}\\
Williamsburg, VA \\
dposhyvanyk@wm.edu}

}

\maketitle

\begin{abstract}
Large Language Models (LLMs) for code are a family of high-parameter, transformer-based neural networks pre-trained on massive datasets of both natural and programming languages. These models are rapidly being employed in commercial AI-based developer tools, such as GitHub CoPilot. However, measuring and explaining their effectiveness on programming tasks is a challenging proposition, given their size and complexity. The methods for \textit{evaluating} and \textit{explaining} LLMs for code are inextricably linked. That is, in order to explain a model's predictions, they must be reliably mapped to fine-grained, understandable concepts. Once this mapping is achieved, new methods for detailed model evaluations are possible. However, most current explainability techniques and evaluation benchmarks focus on model robustness or individual task performance, as opposed to interpreting model predictions. 

To this end, this paper introduces \astxplainer, an explainability method specific to LLMs for code that enables both new methods for LLM evaluation and visualizations of LLM predictions that aid end-users in understanding model predictions. At its core, \astxplainer provides an automated method for aligning token predictions with AST nodes, by extracting and aggregating normalized model logits within AST structures. 
To demonstrate the practical benefit of \astxplainer, we illustrate the insights that our framework can provide by performing an empirical evaluation on 12 popular LLMs for code using a curated dataset of the most popular GitHub projects. Additionally, we perform a user study examining the usefulness of an \astxplainer-derived visualization of model predictions aimed at enabling model users to explain predictions. The results of these studies illustrate the potential for \astxplainer to provide insights into LLM effectiveness, and aid end-users in understanding predictions. 
\end{abstract}

\IEEEpeerreviewmaketitle


\section{Introduction}\label{sec:introduction}

The advent and proliferation of online open-source code repositories and rapid advancements in transformer-based neural large language models (\llms) have served as a catalyst for the advancement of automated Software Engineering (SE) tools with rapidly advancing effectiveness. \llms for code have demonstrated considerable proficiency across a diverse array of generative SE tasks, inclusive of, but not restricted to, code completion \cite{Raychev2014CodeCW,MSR-Completion}, program repair \cite{Chen2019sequencer,ahmad_unified_2021}, and test case generation \cite{Watson:ICSE20}. Moreover, these advancements are rapidly being introduced into commercial developer tools such as GitHub CoPilot\cite{github_copilot} and Replit's Ghostwriter~\cite{ghostwriter}. 

However, the sheer complexity and size that enable the often surprising effectiveness of \llms for code is a double-edged sword. That is, while these attributes enable \llms to capture important patterns in code that allow them to be applied to a range of programming tasks, effectively \textit{explaining} and \textit{evaluating} the capabilities of these models is a challenging proposition --- they effectively function as ``black boxes'' that derive predictions from exceedingly complex internal model mechanics. Current research in both designing \llms for code and in applying them to programming tasks typically makes use of existing benchmarks (\eg CodeSearchNet~\cite{husain2019codesearchnet}, or HumanEval~\cite{chen_evaluating_2021}) and metrics that have been adapted from the field of natural language processing (NLP) such as accuracy, BLEU, METEOR, and ROUGE, as well as more recent metrics further tailored for code such as CodeBLEU~\cite{ren_codebleu_2020}. However, recent work has illustrated the limitations of benchmarks such as HumanEval~\cite{liu2023code}, and there has been growing criticism of automated metrics within the NLP community~\cite{molnar2019interpret,Kim2018InterpretabilityTCAV,wan_what_2022,liu_reliability_2023}. These deficiencies largely stem from the fact that such benchmarks and metrics are often targeted at evaluating functional or syntactic correctness of generated code or task performance, but are not able to \textit{explain} model predictions or capabilities in an interpretable manner. 

Methods for \textit{evaluating} and \textit{explaining} \llms for code are inextricably linked to one another. An informative evaluation requires some degree of explainability of model predictions, such that model behavior can be understood at a fine-grained level. However, the fundamental challenge 
in achieving explainability of \llms for code lies in establishing a reliable mapping mechanism that can bridge the gap between a given model's predictions and human-understandable programming language (PL) concepts that can aid in explaining the model's decisions. As such, designing both effective evaluations and interpretability techniques for \llms of code requires that one first establish this conceptual mapping.

 \begin{figure}
		\centering
		\includegraphics[width=0.5\textwidth]{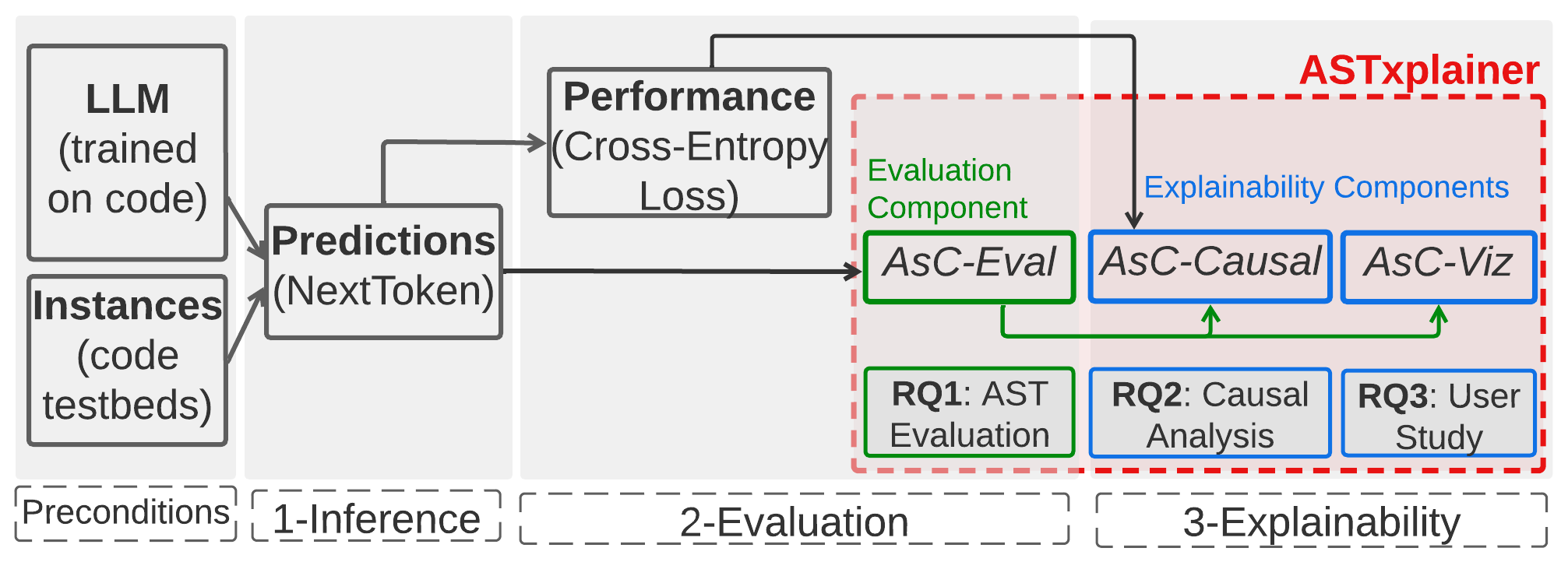}
		\caption{The evaluative and explainability method \astxplainer is composed of \asceval, \asccausal, and \ascviz.}
        \vspace{-0.5cm}
        \label{fig:astxplainer}
\end{figure}

To overcome the challenges in explaining and evaluating \llms for code we propose a novel method for enabling a reliable conceptual mapping of \llm predictions to PL concepts, called \astxplainer, which collects and aggregates \llm token predictions into a construct that we call Abstract Syntax Concepts (\asc), derived from Abstract Syntax Trees (ASTs). By explicitly mapping model predictions to code structure, \astxplainer provides a fine-grained methodology for examining \textit{how} models perform relative to programming language concepts, and can help model end-users reason about \textit{why} an \llm may have made a certain set of predictions. \astxplainer's mapping of model predictions to \ascs enables two new types of evaluations for \llms of code, and one novel interpretability technique that visualizes model \ascs to aid end users (\ie developers using \llms to auto-complete code) in understanding \llm predictions. Fig.~\ref{fig:astxplainer} illustrates these three main components of \astxplainer.

The first evaluation technique, called \asceval is able to estimate the structural performance of a predicted syntax element in order to measure the uncertainty of the downstream code generative process (e.g., for code completion). The second evaluation technique called \asccausal, is capable of generating causal explanations that link these structural performance values with canonical model performance (i.e., Cross-Entropy Loss). Finally, \ascviz implements a practical interpretability technique by visualizing model \llm prediction uncertainty, organized into AST structures, aiding end-users in understanding the reliability of model predictions in practice. We evaluate \asceval and \asccausal through a large-scale, comprehensive empirical study that evaluates 12 popular \llms on a novel dataset of $\approx$ 10 million tokens that are exclusive of the model's training data. Furthermore, to evaluate the effectiveness of \ascviz, we conduct a user study examining the utility of multiple visualizations in aiding developers to understand and explaining model predictions. The results of our empirical study lead to novel insights regarding the performance of \llms for code, and user study illustrates the promising utility of \ascviz. 

The contributions of this paper are as the following: 
\begin{itemize}
    \item An \textit{evaluative metric} based on Abstract Syntax Concepts and Next-token Predictions (\asceval).  
    \item An explainability method (\asccausal) that links canonical evaluations with our Abstract Syntax Concepts to provide insights into why the cross-entropy loss is being affected by structural elements of code data.  
    \item A user study that shows how AST visualizations (\ascviz) help to understand generated code.
    \item A benchmark to evaluate Abstract Syntax Concepts in \llms, which includes a curated dataset (\galeras) of ~50K python samples.
    \item Experimental data, curated datasets, source code, and complementary statistical analysis used in this research are published in an open-source repository, which is available at \url{https://github.com/WM-SEMERU/CodeSyntaxConcept}.  
\end{itemize}


\section{ Background \& Related Work}
\label{sec:background}

\astxplainer is an evaluative and explainability approach to quantify the prediction uncertainty of \llms for code. \llms are the result of scaling up billions of parameters for context-aware word representations from pre-trained models\cite{zhao_survey_2023}. This section defines and formalizes the basic elements of our approach. We provide a definition of \llms and how to evaluate them, the definition of Abstract Syntax Trees (ASTs) and how they were employed for probing, and finally, the explainability methods for \llms. 

\subsection{Large Language Models for Code}

Our research focused on \llms because of their outstanding performance on code-based generative tasks. While other representations exist, such as graph-based models~\cite{allamanis2018learning,Allamanis19}, we focus our discussion on sequence-based representations for simplicity. The goal of sequence-based models is to statistically learn a representation of a software artifact (\eg snippet, comments, or test cases). We refer to SE-specific sequence-based data as a software corpus $\mathcal{S}$. Given the sequential nature of $\mathcal{S}$, we can decompose $\mathcal{S}$ into a desired granularity of tokens, words, or sub-words \cite{Karampatsis2019} by using a transformation function $\Gamma(\mathcal{S})= w_1,...,w_I$ (\ie \textit{tokenizers}). This transformation function is a tokenization method for converting a software corpus into a sequence of discrete objects $w_i$  for $1 \leqslant i \leqslant I$. Note that $w_i \in V$, where the vocabulary $V$ is a finite set.

Given this definition, a statistical language model is a probability distribution $P$ over a fixed granularity of sequences of software corpora $\mathcal{S}$. We can factorize the joint distribution over the $i-$dimension as: $P(\mathcal{S}) = P(w_1,...,w_I) = \prod_{i = 1}^{I} P(w_i | w_{<i})$. Due to the discrete nature of the data, the expression $P(w_i | w_{<i})$ can be estimated using a classifier. The classifier, in our particular case, is a LLM \cite{Bengio2003AModel}. Hence, rather than using \textit{n}-grams or Markov Models to approximate $P(w_i | w_{<i})$ \cite{Karampatsis2020Open-VocabularyAbstract}, it is convenient to use a latent model $P(w_i | w_{<i} ) \approx P(w_i | h_i )$, where $h_i$ is known as a \textit{hidden state} that embeds the sequence information from past observations up to the time step $i$.

Depending on \textit{how} the sequence is processed, the hidden state $h_i$ can be computed using either \textit{Encoder-Only}, \textit{Encoder-Decoder}, or \textit{Decoder-Only} architectures according to the \textit{transformers'} layers ~\cite{vaswani2017transformers}. One popular bidirectional objective function used widely in representation learning is \textit{masked language} modeling \cite{devlin_bert_2019}. This function aims to predict masked text pieces based on the surrounding context. CodeBERT \cite{feng_codebert_2020}, CuBERT (345M) \cite{kanade_learning_2020} CodeRoBERTa  \cite{lin_span_2022}, and GraphCodeBERT \cite{guo_graphcodebert_2021} are examples of \textit{Encoder-Only} models for code. In programming contexts, these methods provide useful representations of code sequences for downstream tasks such as code classification, clone and defect detection. CodeT5 \cite{wang_codet5_2021} and PLBART \cite{ahmad_unified_2021} are examples of \textit{Encoder-Decoder} models. These models encode an input sequence and, then, this encoded sequence is decoded with a different architecture. Encoder-Decoder models are trained with the goal of reconstructing masked input sequences \cite{lewis_bart_2019}. Additionally, they have been employed for SE tasks such as code summarization, and code generation using masks\cite{wang_codet5_2021}. Finally, \textit{Decoder-Only} models predict the probability of a token given a preceding sequence. CodeGPT \cite{lu_codexglue_2021}, CodeParrot \cite{codeparrot}, GPT-Neo \cite{black_sid_2021_5297715}, GPT-J \cite{gpt-j}, Codex \cite{openai_codex}, GPT-NeoX \cite{GPTNeoX}, and Google's left-to-right decoder-only Transformer language models \cite{vaswani2017transformers,austin2021program} are examples of \textit{Decoder-Only} models for code. 

Although our proposed approach \astxplainer was designed to be compatible with either type of \llms, this paper concentrated on \textit{Decoder-Only} models due to their popularity for code-based generative tasks \cite{xu_systematic_2022}. These models share a common property: \textit{the ability to connect previously processed information to a present task, such as using an initial sequence of tokens to predict new code tokens}. The resulting auto-completed sequence should be coherent with respect to the context of the initial sequence. This property is known as the ability to model \textit{long-range dependencies}~\cite{karpathy2015understand}. 

\begin{definition}
\label{def:llm}
  \textbf{Decoder-Only Transformers.} Decoder-Only models update the hidden state $h_i = f(h_{i-1}, w_{<i})$ using past inputs $w_{<i}$ and a previous hidden state $h_{i-1}$. In other words, these models function in a feed-forward manner that predicts future values from historical values directly. \llms trained on source code have the ability to generate tokens or sub-words given a history. Hence, decoder-only models are employed as generative models $\hat{w_i} \backsim P(w_i | w_{<i} ) = \sigma(y)_i = \frac{e^{y_{w_i}}}{\Sigma_j e^{y_j}}$. 
\end{definition}

\noindent In the previous approximation, the predicted token $w_i$ is \textit{conditioned} by the previous information. The term $y_j$ represents the \textit{non-normalized log-probabilities} for each output token $j$. We extracted and normalized these \textbf{log-probabilities} from the last layer of \llms to estimate the \textbf{Next-token Predictions} (\ntp) in \astxplainer (see Sec.\ref{sec:approach}). This estimation relies on the softmax function. The softmax $\sigma_i$ returns a distribution over predicted output classes, in this case, the classes are each token in the previously introduced vocabulary $V$. It is expected that the predictions contained in $\sigma_i$ are influenced by previous inputs of the sequence $w_{<i}$.

\subsection{ASTs Probing Approaches}

\textit{Probing} is a supervised analysis to determine which type of parameters (\eg input code snippets, tokenization process, number of hidden layers, and model size) influence the learning process in machine learning models \cite{troshin_probing_2022}. The purpose of probing is to assess whether hidden representations of machine learning models (\ie \llms) encode specific linguistic properties such as syntactic structures of programming languages. For example, Lopez \etal \cite{lopez_ast-probe_2022} trained a linear classifier to show that code syntactic structures are encoded in pre-trained models in the form of Abstract Syntax Trees (ASTs). Lopez \etal's approach demonstrates that the middle layers of pre-trained models contain ASTs' information\cite{lopez_ast-probe_2022}.

Nonetheless, instead of proposing another syntax probe, our approach \astxplainer adapts AST information to evaluate and explain \llms (see Sec.~\ref{sec:approach}). ASTs are defined as a formal representation of syntactical structures built upon linguistic elements of PLs. ASTs are formed according to the production rules defined in Context Free Grammar (CFGs). More precisely, production rules are functions that combine terminal and non-terminal nodes into statements. Terminal nodes are symbols in the source code (\eg tokens in region \circled{3} of ~Fig.\ref{fig:local_evaluation}), while non-terminal nodes encapsulate more than one terminal node to define the structure of a statement (\eg nodes containing children in region \circled{2} of ~Fig. \ref{fig:local_evaluation}). 

When designing our approach \astxplainer (see Sec.\ref{sec:approach}), we leveraged meaningful and interpretable information defined in Context-Free Grammars ($CFGs$). $CFGs$ are a set of rules containing the syntax and structural information of a language \cite{10.5555/1196416}. Ultimately CFGs define instructions that specify how different tokens (\ie Lexemes) are put together to form valid statements in every programming language.

\begin{definition}
\label{def:cfg}
\textbf{Context Free Grammars.} $CFG$ $\mathbb{G}$ is expressed as $\mathbb{G} = (\alpha, \lambda, \omega, \beta)$ where $\alpha$ denotes the finite set of non-terminal symbols, $\lambda$ the finite set of terminal symbols, $\omega$ the finite set of production rules and $\beta$ the start symbol. The set of production rules $\omega$ for any type of statement (\eg conditional, assignation, operator) is expressed in terms of the terminal and non-terminal symbols.
\end{definition}

\subsection{Explainability for Code Generation}

\llms for code can be considered a black box because of their uncertain behavior when predicting tokens. To estimate such uncertainty, we can employ \textit{explainability} methods on \llms. Explainability aims to understand how a model operates and comes to decisions either by exploring inner layers or performing perturbation analysis on the models' inputs \cite{belle_principles_2020,molnar_interpretable_2020}. For example, Gholizadeh \etal \cite{gholizadeh_model_2021} propose a local explainability technique, namely layer-wise relevant propagation (LRP), that computes the importance of an interpretable \textit{n}-gram in classifying a text sequence. LRP calculates a score with the sum of activated weights during the back-propagation to identify the most influential \textit{n}-grams. This score is employed for explaining the importance of a given \textit{n}-gram for a canonical (\ie SVM) and a neural model(\ie CNN). The authors demonstrated that LRP outperforms the gradient-only-based and permutation-only-based explainability techniques \cite{gholizadeh_model_2021}. It is important to clarify that, in our research, \textit{explainability} and \textit{interpretability} are used interchangeably. 

In the context of pre-trained models for code, Liu \etal experimented with Encoder-Decoder models for code2code and comment2code tasks (\eg T5, CodeText, and CodeTrans). Their research aims at explaining why neural models generate code sequences reliably by identifying tokens that contribute the most to a sequence prediction \cite{liu_reliability_2023}. Moreover, Vasconcelos \etal propose a technique that highlights generated code using an uncertainty threshold. Their approach points out fragments of the sequence where developers can intervene upon the uncertainty threshold \cite{vasconcelos_generation_2023}. On the other hand, we can explain pre-trained models for code using structural information. For instance, Wan \etal conducted an interpretability analysis on Encoder-only models (\eg CodeBert and GraphCodeBert) focusing on three aspects: 1) how the self-attention weights align with the syntax structure, 2) whether the syntax structure is encoded in the hidden layers, and 3) how pre-trained models induce syntax structure \cite{wan_what_2022}. 

Even though previous research has introduced explainability techniques to analyze pre-trained models with structural information, those techniques  have been tested and designed for modest-size Encoder-Only models (\ie less than 1B). Conversely, our study \astxplainer proposes not only an explainability technique that contextualizes canonical metrics (\ie cross-entropy loss) based on causal inference (see Fig.\ref{fig:asc_causal}) but also an evaluative metric (\asceval) for Decoder-only \llms that predicts ASTs terminal and non-terminal nodes. More importantly, we introduce and control a set of confounders based on code features (\eg AST-levels, AST-nodes, and number of tokens) to properly estimate the relationship between \asceval and canonical metrics (see Tab.~\ref{tab:correlations}).

Kim \etal \cite{Kim2018InterpretabilityTCAV} introduce a formal mathematical structure known as a \textbf{function for explainability} ($\varphi$). We use this definition to formally describe what constitutes an explainable method in SE. Most \llms for code operate by predicting tokens $P(w_i|d_i)$ that do not \textit{inherently} match high-level concepts a human can easily understand. Kim \etal claim that such difficulty can be expressed mathematically as representing the state of \llms as a vector space ($\Vec{m}$). Conversely, humans or, in our study, developers operate in a different vector space $\vec{h}$, which corresponds to an unknown set of \textbf{human-interpretable concepts} ($h$). As such, our main challenge is to map $\Vec{m} \to \Vec{h}$ bridging this gap between the disparate vector spaces. The \textit{key insight} of \astxplainer is the formalization of an explainability function $\varphi$ for \llms of code.

\begin{definition}
\label{def:explainability_function}
\textbf{Interpretability Function for Next Token Predictions.} Consider $\varphi: \Vec{m} \to \Vec{h}$. In this formulation, $\Vec{m}$ represents an approximation of a model's vector space as measured through token prediction performance at different granularity levels (\ie normalized log-probabilities). This vector space approximation is then mapped to human-understandable concepts $\Vec{h}$ that represent programming language syntactic concepts (\ie terminal and non-terminal nodes).
\end{definition}
\section{The \asceval Component}
\label{sec:approach}

\begin{figure*}[ht]
		\centering
		\includegraphics[width=1\textwidth]{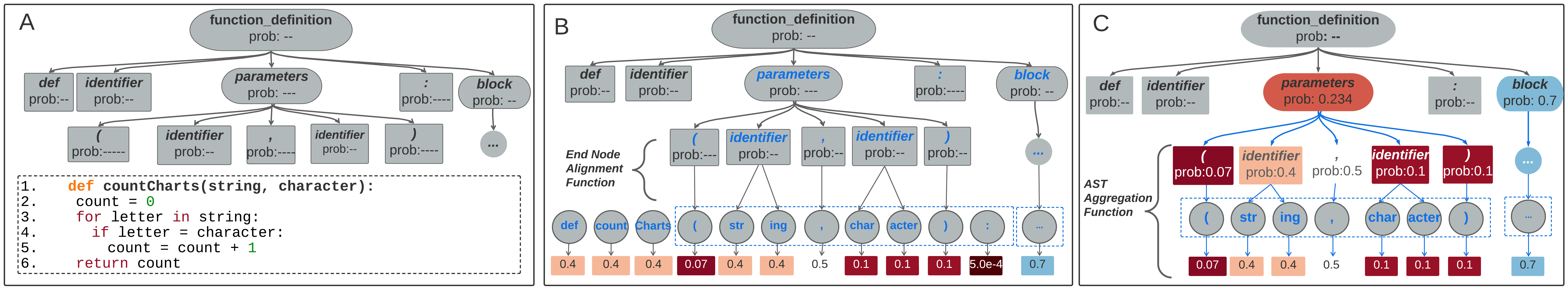}
		\caption{\asceval Components. Left: Nodes are employed as ``concepts''. Center: Each token is aligned to the end nodes of the AST with an offset function. Right: Node probabilities are estimated with an aggregation function.}
        \vspace{-1em}
        \label{fig:ast_eval}
\end{figure*}

While LLMs have seen striking advances with regard to code generation and other downstream SE tasks \cite{Chen2021EvaluatingCode, watson2020dl4se}, researchers are still not able to evaluate what aspects of code are actually statistically learned by these models. In this section, we propose a new metric, \asceval, to showcase the statistical behavior of syntactic elements generated by \llms. Our proposed \asceval comprises the basic units for explainability (see Fig.~\ref{fig:ast_eval}) as Abstract Syntax Concepts (\asc), an alignment function $\delta$ that links tokens with ASTs, and an aggregation function $\theta$ that estimates the prediction performance of a terminal and non-terminal nodes. We propose an explainability function $\varphi$ that relies on the alignment function $\delta$ and the aggregation function $\theta$ to perform the mapping from log-probabilites (\ie \ntp) to developer-understandable concepts (\ie \asc).

\subsection{Abstract Syntax Concepts (\asc)}

\asceval can be formally defined (see Def.~\ref{def:explainability_function}) as an explainability function $\varphi$ of token predictions of \llms using Context Free Grammars. We introduce the term \textbf{Abstract Syntax Concepts} (\asc) to represent the terminal and non-terminal symbols in a Context Free Grammar (see Def~.\ref{def:cfg}). Specifically, to approximate a \llms' vector space, in $\Vec{m}$, we extract the last layer to calculate \ntp, which is, in fact, a generative measure of performance. Then in $\Vec{h}$, we map the model's prediction performance at the token level (\ntp) to \asc (for which we define a set of categories $\mathcal{H}$), to make it easier to interpret what aspects of \llms are \textit{effective} or \textit{erroneous} at predicting. 

In PLs, terminal and non-terminal nodes retain different semantic meanings. For instance, \texttt{\small `identifier'} and \texttt{\small `string'} nodes correspond to a common \textit{Natural Language} concept category. As such, we can group nodes $n$ into semantically meaningful \textit{categories} $\mathcal{H}$. Fig.~\ref{fig:largeTreeMap} depicts some of our proposed categories for Python. These categories will allow \asceval to assign semantic meaning to predicted \asc. \asc are the fundamental mathematical units for enabling the evaluation and explainability of \llms. Figure \ref{fig:largeTreeMap} depicts some of the concepts used to evaluate \llms with \asceval. Concepts $n \in N$ are types of symbols defined by tree-sitter's $CFG$ \cite{tree-sitter}. In summary, Each token in a sequence $s$ can be assigned to a category $h \in \mathcal{H}$. With our categories $\mathcal{H}$, researchers and developers can easily associate \llms' performance to particular structural code attributes. As such, \asceval allows for \llms Next-token Predictions to be explained in a developer-centric way.

Fig~\ref{fig:ast_eval}-A depicts the AST representation of a Python snippet of a naive implementation of the function $countCharts$. This function counts and returns the number of occurrences of a given character for an input string. In the AST representation, the leaf nodes correspond to the terminal tokens used in the snippet, while the intermediate nodes correspond to non-terminals. Our approach relies on the tree-sitter library \cite{tree-sitter} to construct the AST representations of the snippets. Once the AST has been parsed, we can access the information for all nodes and retrieve useful properties such as their type, children, and location.

\subsection{AST Alignment function ($\delta$)} Figure~\ref{fig:ast_eval}-B illustrates the process of aligning terminal and non-terminal nodes in the AST representation with their corresponding tokens. Prior to this alignment process, we split the $countCharts$ snippet $s$ into tokens using the model tokenizer $\Gamma(s) = (w_1,...,w_i)$. Since the tokenizer may produce a sequence of tokens where each token does not necessarily matches with a single terminal node, a single node in the AST may contain more than one associated token. In fact, intermediate nodes are aligned with a sub-sequence of the original snippet rather than a single token. We define for this purpose the alignment function $\delta: N \to s_{<=i}$ where $s_{<=i}$ corresponds to a subsequence of a snippet and $N$ is the set of terminal and non-terminal nodes. We leverage the offset property of each AST node to conduct this process, in other words, we search for all the tokens in $s$ that are located within the offset range of each node. To illustrate how function $\delta$ works, let's consider the example in Figure~\ref{fig:ast_eval}-B, in the sub-tree the terminal node \texttt{\small `('} is aligned with token \framebox(0.5cm,0.34cm){(} while the sibling node \texttt{\small `identifier'} is aligned with tokens \framebox(0.7cm,0.34cm){str} \framebox(0.7cm,0.34cm){ing}. The parent node \texttt{\small `parameters'} will be consequently aligned with \framebox(0.5cm,0.34cm){(} \framebox(0.7cm,0.34cm){str} \framebox(0.7cm,0.34cm){ing} \framebox(0.5cm,0.34cm){,} \framebox(0.8cm,0.34cm){char} \framebox(0.9cm,0.34cm){acter} \framebox(0.5cm,0.34cm){)}.

\subsection{AST Aggregation function ($\theta$)}
We design an aggregation function $\theta$ that computes our proposed metric \asceval, which represents how confident a terminal or non-terminal node $n$ is predicted by an \llm. By relating these node predictions to an actual node symbol, we gain an understanding of how well a studied model is \textit{generating code}. These \asceval performance values can also uncover specific long-range interactions and map them into an AST visual structure (see Sec.~\ref{sec:approach-3}). \asceval performs at two levels of granularity depending on the scope of the analyzed corpus $\mathcal{S}$. We refer to such granularity as \textit{local} and \textit{global} aggregation. Local aggregations operate for a code snippet, while global aggregations operate for a corpus. Although local aggregation can provide a \asceval value for a single snippet, this aggregation allows computing an average of aggregated values at snippet granularity.  

Figure~\ref{fig:ast_eval}-C shows the aggregation function used to compute the prediction probability for each node. Once the tokens are aligned with their corresponding nodes using $\delta$, we traverse the entire AST and aggregate the \ntp probabilities of their associated tokens. The aggregation function $\theta$ can take the form of a statistical average, median or max values depending on the user configuration. In our study, we set the aggregation $\theta: N \to  median(\delta(N))$ for a subset of tokens $s_{<=i}$. For example, as illustrated in Fig.~\ref{fig:ast_eval}-C, the parent node \texttt{\small `parameters'} has an associated average value of $0.23$. This parent node average was aggregated with its terminal values: \texttt{\small `('} with $0.07$, \texttt{\small `identifier'} with $0.4$, \texttt{\small `,'} with $0.5$, \texttt{\small `identifier'} with $0.1$, and \texttt{\small `)'} with $0.1$.

\begin{figure}[h]
\centering
\includegraphics[width=0.47\textwidth]{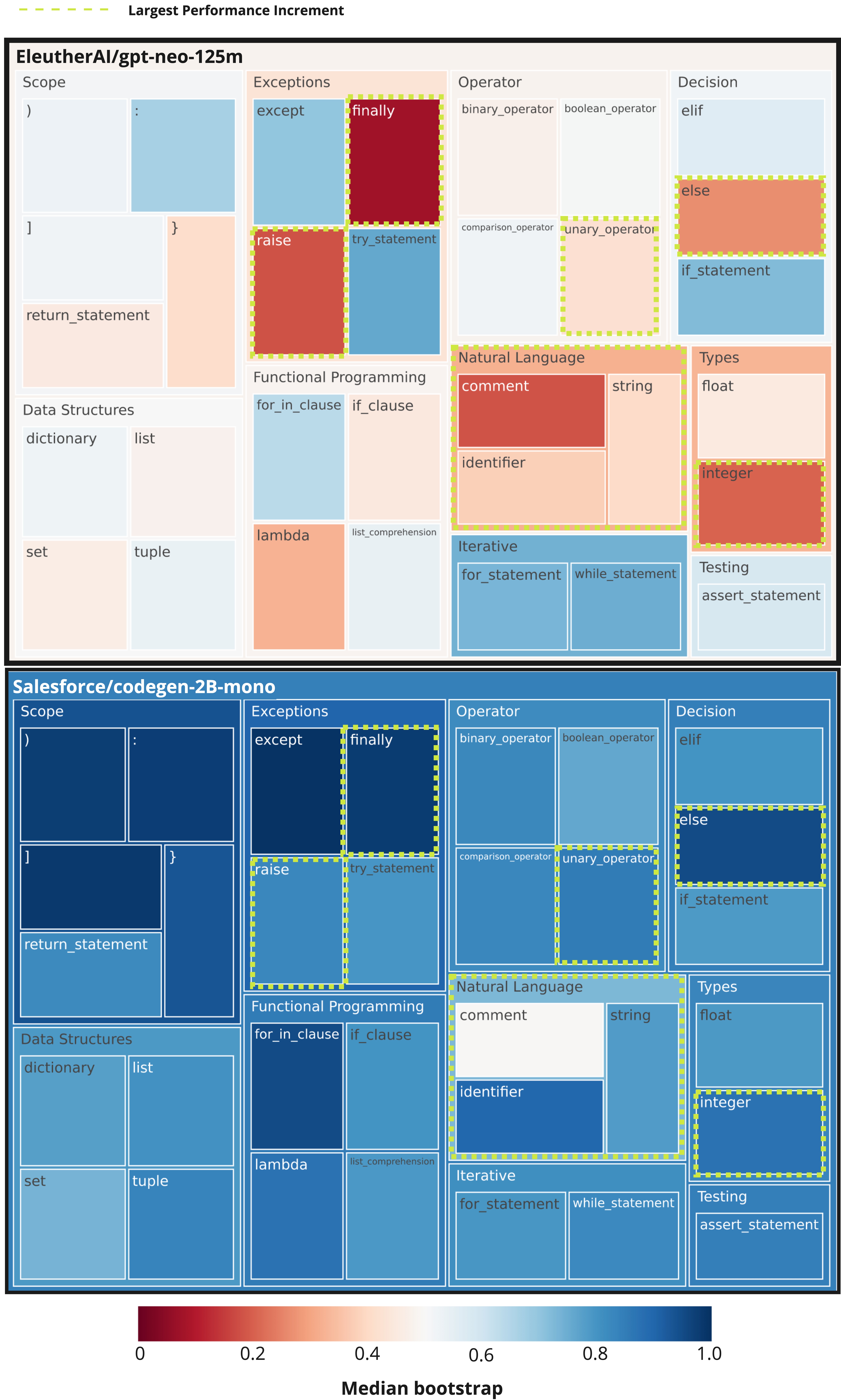}
\caption{\asceval for 10 \asc Categories and 2 \llms (\monoIIII and \gptI)}
\vspace{-1em}
\label{fig:largeTreeMap}        
\end{figure}

\section{The \asccausal Component}
\label{sec:approach-2}
In this section, we show how \asccausal component can be used to explain and contextualize other canonical metrics such as the cross-entropy loss. To achieve that, we propose a causal inference technique to estimate the impact of Abstract Syntax Concepts (\asc) predictions on overall \llm performance. 

\begin{figure}[ht]
		\centering
		\vspace{-1.5em}
  \includegraphics[width=0.45\textwidth]{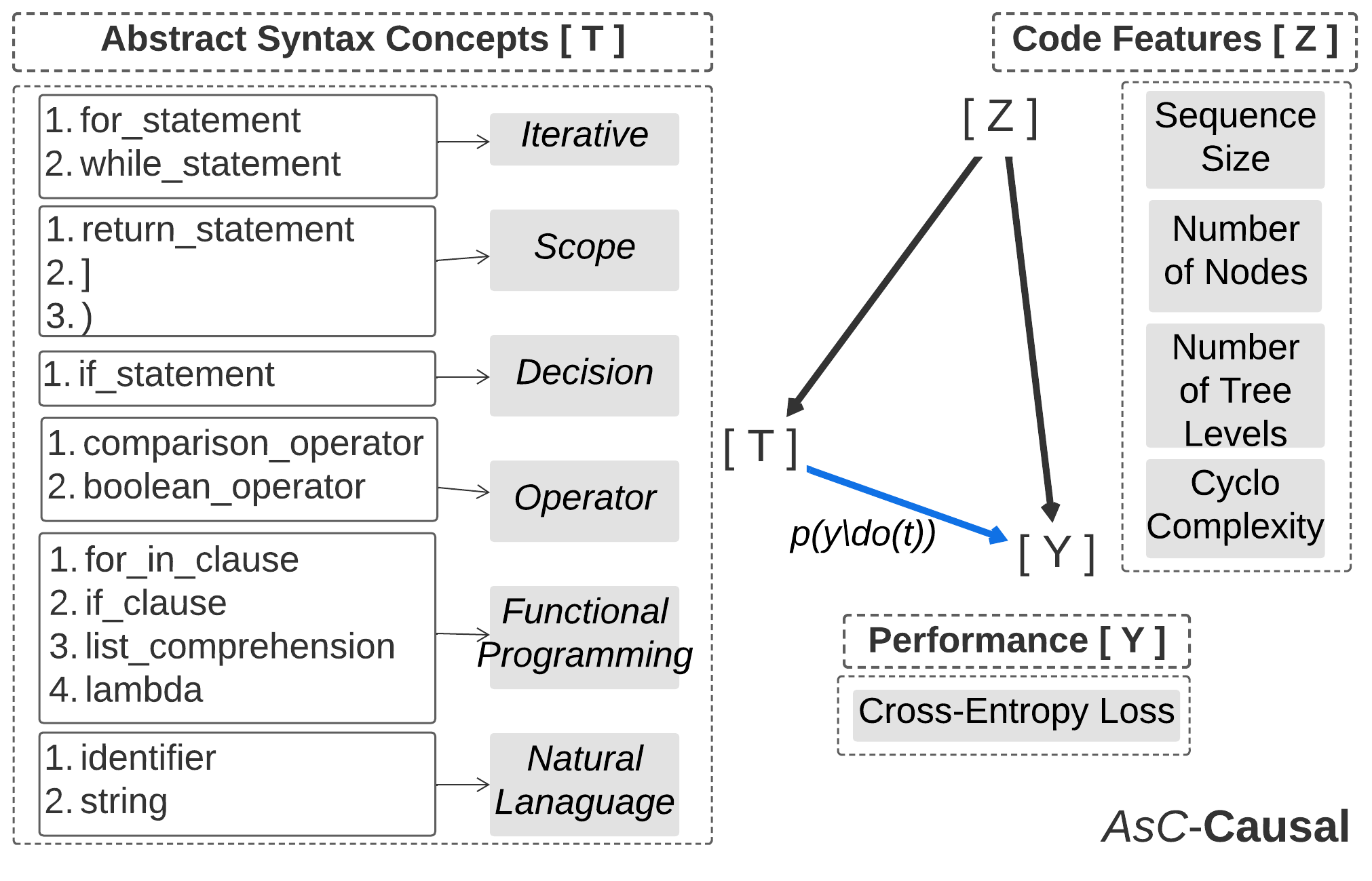}
		\caption{\textbf{Structural Causal Model} to estimate the Average Treatment Effect of \asc to \llm Performance by controlling code features (\asccausal).}
            \vspace{-1.5em}
        \label{fig:asc_causal}
\end{figure}

\llms are more understandable when they \textit{reflect human knowledge} \cite{Kim2018InterpretabilityTCAV}. One way of determining whether an \llm for code reflects human knowledge is testing it to see whether or not it operates \textit{similar to how a developer would estimate the prediction of a sequence}\cite{palacio2023theory}. For instance, consider the situation where a developer inserts a \texttt{\small `for statement'} in a snippet. Inherently, a developer mentally rationalizes several things such as the concept of \textit{Iteration}. If an \llm is able to make a similar prediction, it suggests to us that it has \textit{statistically learned} some understanding of the syntax structure of a programming cycle. We can consider that this statistical behavior impacts the cross-entropy loss. This impact indicates that Abstract Syntax Concepts (\asc) are influencing the quality of a \llm (see Def.~\ref{def:effect}). In order to estimate such influence, we propose a causal inference technique based on the do-calculus analysis \cite{Pearl2016Causality,Pearl2009Causality,Pearl2018Causality}. For instance, in Eq.~\ref{eqn:do-all-lines}, we compute a causal effect (Eq.~\ref{eqn:do-1}) and correlation (Eq.~\ref{eqn:do-1}) for the concept treatment \texttt{\small `for\_statement'} impacting the cross-entropy loss of a given \llm.

\begin{subequations}
\begin{align}
p(Y|do(t=forState.)) &=\sum_{z \in _{codeFeat.}}p(Y|z,t)p(t)\label{eqn:do-1} \\
p(Y|t=forState.) &= \sum_{z \in _{codeFeat.}}p(Y|z,t)p(t|z) \label{eqn:do-2} 
\end{align}
\label{eqn:do-all-lines}
\end{subequations}

We can explain the prediction performance of \llms using \asceval values as treatment effects. These effects are computed from a \textbf{Structural Causal Model} (SCM), which represents our assumptions about the underlying causal process. In our study, these assumptions take the form of the performance of each \asc  (treatments $T$), code features (confounders $Z$), and the \llms canonical performance (outcome $Y$). The relationship or directionality information of these causal variables is explicitly stated in the SCM (see Fig.~\ref{fig:asc_causal}). The goal of the causal analysis is to determine the \textit{Average Treatment Effect} (ATE) that a treatment has on the outcomes after controlling the confounding variables. In other words, we want to estimate the probability $p(Y|do(T))$ (see Eq.~\ref{eqn:do-1}) to identify cases of \textit{spurious correlations} (\ie association is not causation) \cite{Sharma2021DoWhyAssumptions}. Note that the probability  $p(Y|do(T))$ is different from $p(Y|T)$ in Eq.~\ref{eqn:do-2}. We state that the probability  $p(Y|T)$ represents the correlation between the variables $Y$ and $T$ without controlling any confounder's effects on treatments or outcomes. In our study, we compute the Pearson correlation $\rho = p(Y|T)$. Conversely, the treatment effect $p(Y|do(T))$ is estimated with \textit{a liner regression} after applying the \textit{the back-door criterion} for controlling confounders \cite{Sharma2021DoWhyAssumptions}. 

\begin{definition}\label{def:effect}
\textbf{\asc Causal Treatment Effects.} Given a Structural Causal Model where a set of variables $PA$ denotes the parents (\ie code features) of $T$, the treatment effect of T (\ie \asc) on Y (\ie cross-entropy loss) is given by
\begin{subequations}
    \begin{align}
    p(Y=y|do(T=t)) &=  \label{eq:effect-1}\\
    \Sigma_z p(Y=y|T=t,PA=z)p(PA=z)  &= \label{eq:effect-2}\\
    \Sigma_z p(T=t,Y=y,PA=z)/p(T=t|PA=z) \label{eq:effect-3}
    \end{align}
\label{eq:effect}
\end{subequations} 
  
\end{definition}

Based on the causal inference definition by Pearl \etal \cite{Pearl2016Causality}, we propose a specific treatment effect for our Abstract Syntax Concept \asc. Def.~\ref{eq:effect} depicts the statistical marginalization of confounders. In simple terms, the Average Treatment Effect comprises the \textit{pure} impact of the treatment in the outcome without the influence of confounding variables. These effects represent the slope of the linear model obtained between the treatment and the output after controlling for confounding.  In our study, we controlled for confounders such as \textit{sequence size}, \textit{number of nodes}, \textit{number of tree levels}, and \textit{cyclomatic complexity}. 
\section{The \ascviz Component}
\label{sec:approach-3}

The visualization component \ascviz is a graphical explainability technique that displays the \asceval performance values of the terminal and non-terminal nodes for a \textit{single} local evaluation. We take advantage of the hierarchical structure of PLs to visually accommodate \asceval values obtained in \asceval into the AST. Fig. \ref{fig:local_evaluation} illustrates how we accommodate the \asceval values for a code generation task using our analyzed \gptII model. Region \circled{1} shows a box with a prompt with an incomplete snippet followed by a second box with generated tokens in blue. Then, in region \circled{2}, the resulting auto-completed snippet is processed with \asceval and represented as an AST. Each node has information about the \asceval performance after applying local aggregations $\theta$. The nodes are color-coded. The highest aggregated values (\ie best predictions) are displayed in shades of blue. In contrast, nodes with the smallest values (\ie worst predictions) are displayed in shades of red. Nodes, in region \circled{2}, encapsulate the code tokens generated by the \llm as presented in region \circled{3}. We refer to tokens linearly organized as \textit{sequence representation}.  

\begin{figure*}[ht]
		\centering
            \vspace{-2em}
		\includegraphics[width=0.9\textwidth]{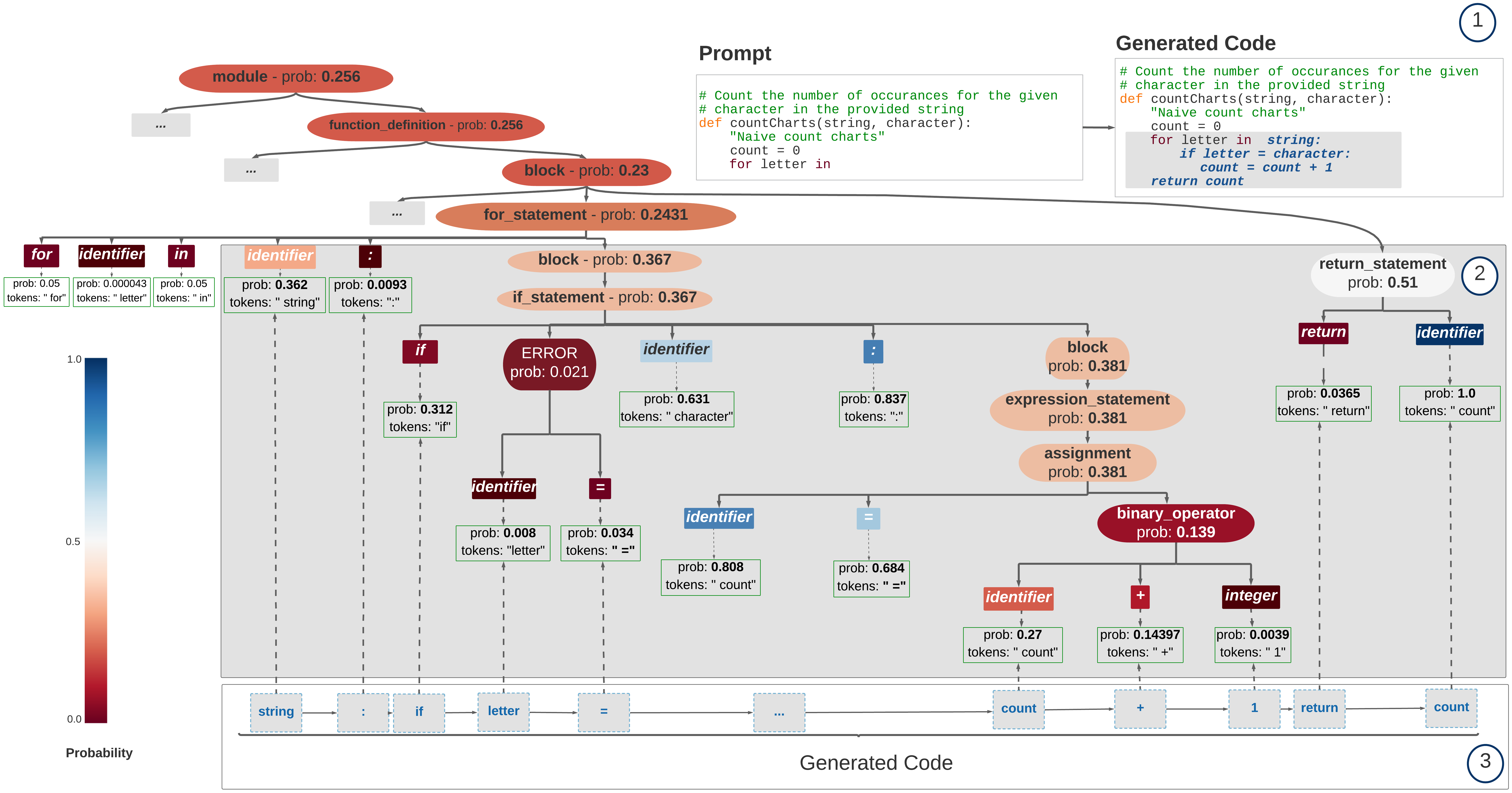}
		\caption{Local Evaluation for Code Completion (\ascviz).}
            \vspace{-1em}
        \label{fig:local_evaluation}
\end{figure*}

\section{Experimental Design}
\label{sec:design}

In order to illustrate the insights that \astxplainer can enable, we present an empirical evaluation on 12 \llms, which shows how \llms behave for each Abstract Syntax Concept, and a user study, which assesses the usability of our approach. This section details the methodological steps we followed to configure, evaluate, and explain our selected \llms. 

\begin{enumerate}[label=\textbf{RQ$_{\arabic*}$}, ref=\textbf{RQ$_{\arabic*}$}, wide, labelindent=5pt]\setlength{\itemsep}{0.2em}
      \item \label{rq:asteval} { \textbf{\asceval:} \textit{To what extent do Large Language Models for code predict syntactic structures?}} 
      \item \label{rq:astxplainer} {\textbf{\asccausal:} \textit{How do Abstract Syntax Concepts impact LLMs' canonical prediction performance?}}
      \item \label{rq:usability} {\textbf{\ascviz:} \textit{How useful is our AST evaluation method for developers in a practical scenario?}}
\end{enumerate}

\subsection{Study Setup} 

\textit{Data Collection.} Our selected \llms were trained on \textit{BigQuery} \cite{bigquery}, \textit{BigPython}\cite{icodegen}, and the \textit{Pile} \cite{gao2020pile}. These datasets include repositories and files from GitHub created before 2021. However, in order to properly evaluate \asceval, we must avoid data contamination. That is, we need to avoid using samples in the evaluation that \llms have already been used for training. For the same reason, we cannot evaluate our approach using popular code datasets such as \textit{CodesearchNet}\cite{husain2019codesearchnet} or \textit{CodeXglue}\cite{lu2021codexglue}. To solve this data contamination issue, we collected ~50k unique Python snippets and created a brand new code dataset, called \galeras. \galeras contains only recent commits performed from January 1st, 2022 to January 1st, 2023. We collected Python repositories from Github that have more than one hundred stars and extracted snippets of code from new and updated Python methods. We cleaned sample duplicates using the commits' history. Additionally, \galeras includes information about the commit message, comments on the method, the whole AST data structure of the method, number of nodes, AST levels, AST errors, white spaces, lines of code, cyclomatic complexity, and token counts.

\textit{Model Collection.} We evaluated and explained a total of 12 open Decoder-Only \llms filtered by popularity. Our largest model has 2.7B parameters.  Table~\ref{tab:models} shows the \llms grouped into four different categories that correspond with the fine-tuning strategy employed. The first category consists of GPT-3-based models trained mostly on natural language (\ie Pile\cite{gao2020pile}). The second category includes models trained on natural language but built upon the \textit{codegen} architecture \cite{nijkamp2023codegen}. The third category consists of models trained on multiple programming languages using BigQuery\cite{bigquery} on both gpt-2 and codegen architectures. The last category corresponds to both \multilang models  fine-tuned on BigPython\cite{nijkamp2023codegen}, which we refer to as \monolang, and gpt-2 models (\ie codeparrot \cite{codeparrot}).

\textit{Machine Configuration.} We performed the experiments using 20.04 Ubuntu with an AMD EPYC 7532 32-Core CPU, A100 NVIDA GPU with 40GB VRAM, and 1TB RAM. For the model inference process, we used HugginFace and Pytorch \cite{wolf2020transformers, pytorch}. All models were loaded into the GPU of the machine to boost the inference time. 

\begin{table*}[h]
\centering
\caption{Large Language Models characteristics and their associated \asceval performance. Erroneous \asceval values are in red. Confident \asceval values are in blue. Best global \asceval is underlined.}
\vspace{-0.2cm}
\label{tab:models}

\scalebox{0.85}{%
\setlength{\tabcolsep}{5pt} 


\begin{tabular}{@{}llll|ccccccccccc@{}}
\toprule
\multicolumn{4}{c|}{\textbf{Large Language Models (LLMs)}} &
  \multicolumn{11}{c}{\textit{\textbf{\asceval Performance (bootstrapped median)}}} \\ \midrule
\multicolumn{1}{c}{\textit{Type}} &
  \multicolumn{1}{c}{\textit{Name}} &
  \multicolumn{1}{c}{\textit{Architecture}} &
  \multicolumn{1}{c|}{\textit{Size}} &
  \multicolumn{1}{c|}{\textit{\textbf{Global}}} &
  \multicolumn{1}{l}{Data Str.} &
  \multicolumn{1}{l}{Decision} &
  \multicolumn{1}{l}{Except.} &
  \multicolumn{1}{l}{F. Prog.} &
  \multicolumn{1}{l}{Iter.} &
  \multicolumn{1}{l}{NL} &
  \multicolumn{1}{l}{Oper.} &
  \multicolumn{1}{l}{Scope} &
  \multicolumn{1}{l}{Testing} &
  \multicolumn{1}{l}{Types} \\ \midrule
 &
  gpt-neo-125m\cite{black_sid_2021_5297715} &
  \textit{gpt-3} &
  125M &
  \multicolumn{1}{c|}{0.48} &
  0.50 &
  0.52 &
  \cellcolor[HTML]{FFCCC9}{\color[HTML]{000000} \textbf{0.43}} &
  \cellcolor[HTML]{FFCCC9}{\color[HTML]{000000} \textbf{0.49}} &
  0.74 &
  \cellcolor[HTML]{FFCCC9}{\color[HTML]{000000} \textbf{0.32}} &
  \cellcolor[HTML]{FFCCC9}{\color[HTML]{000000} \textbf{0.48}} &
  0.51 &
  0.59 &
  \cellcolor[HTML]{FFCCC9}{\color[HTML]{000000} \textbf{0.33}} \\
 &
  gpt-neo-1.3B\cite{black_sid_2021_5297715} &
  \textit{gpt-3} &
  1.3B &
  \multicolumn{1}{c|}{0.59} &
  0.60 &
  0.61 &
  0.53 &
  0.62 &
  0.79 &
  \cellcolor[HTML]{FFCCC9}{\color[HTML]{000000} \textbf{0.43}} &
  0.57 &
  0.68 &
  0.68 &
  \cellcolor[HTML]{FFCCC9}{\color[HTML]{000000} \textbf{0.44}} \\
\multirow{-3}{*}{\textbf{\begin{tabular}[c]{@{}l@{}}Natural L.\\ gpt-3\end{tabular}}} &
  gpt-neo-2.7B\cite{black_sid_2021_5297715} &
  \textit{gpt-3} &
  2.7B &
  \multicolumn{1}{c|}{{\ul \textbf{0.62}}} &
  0.62 &
  0.63 &
  0.56 &
  0.66 &
  \cellcolor[HTML]{CACDFA}\textbf{0.81} &
  \cellcolor[HTML]{FFCCC9}{\color[HTML]{000000} \textbf{0.46}} &
  0.60 &
  0.74 &
  0.70 &
  \cellcolor[HTML]{FFCCC9}{\color[HTML]{000000} \textbf{0.47}} \\ \midrule
 &
  codegen-350M-nl\cite{Nijkamp2022ACP}&
  \textit{codegen} &
  350M &
  \multicolumn{1}{c|}{0.55} &
  0.56 &
  0.57 &
  \cellcolor[HTML]{FFCCC9}{\color[HTML]{000000} \textbf{0.45}} &
  0.57 &
  0.77 &
  \cellcolor[HTML]{FFCCC9}{\color[HTML]{000000} \textbf{0.39}} &
  0.54 &
  0.64 &
  0.64 &
  \cellcolor[HTML]{FFCCC9}{\color[HTML]{000000} \textbf{0.40}} \\
\multirow{-2}{*}{\textbf{\begin{tabular}[c]{@{}l@{}}Natural L.\\ codegen\end{tabular}}} &
  codegen-2B-nl\cite{Nijkamp2022ACP}&
  \textit{codegen} &
  2B &
  \multicolumn{1}{c|}{{\ul \textbf{0.65}}} &
  0.65 &
  0.65 &
  0.58 &
  0.68 &
  \cellcolor[HTML]{CACDFA}\textbf{0.82} &
  \cellcolor[HTML]{FFCCC9}{\color[HTML]{000000} \textbf{0.48}} &
  0.61 &
  0.78 &
  0.72 &
  0.50 \\ \midrule
 &
  codeparrot-small-multi\cite{codeparrot} &
  \textit{gpt-2} &
  110M &
  \multicolumn{1}{c|}{0.57} &
  0.54 &
  0.55 &
  0.64 &
  0.60 &
  0.60 &
  \cellcolor[HTML]{FFCCC9}{\color[HTML]{000000} \textbf{0.40}} &
  0.54 &
  0.71 &
  0.67 &
  \cellcolor[HTML]{FFCCC9}{\color[HTML]{000000} \textbf{0.42}} \\
 &
  codegen-350M-multi\cite{Nijkamp2022ACP}&
  \textit{codegen-350M-nl} &
  350M &
  \multicolumn{1}{c|}{0.68} &
  0.63 &
  0.72 &
  0.75 &
  0.70 &
  0.69 &
  0.51 &
  0.62 &
  0.83 &
  0.73 &
  0.51 \\
\multirow{-3}{*}{\textbf{\begin{tabular}[c]{@{}l@{}}Multi-\\ Language\end{tabular}}} &
  codegen-2B-multi \cite{Nijkamp2022ACP}&
  \textit{codegen-2B-nl} &
  2B &
  \multicolumn{1}{c|}{{\ul \textbf{0.79}}} &
  0.74 &
  0.79 &
  \cellcolor[HTML]{CACDFA}\textbf{0.83} &
  0.81 &
  0.77 &
  0.65 &
  0.74 &
  \cellcolor[HTML]{CACDFA}\textbf{0.91} &
  0.80 &
  0.71 \\ \midrule
 &
  codeparrot-small\cite{codeparrot} &
  \textit{gpt-2} &
  110M &
  \multicolumn{1}{c|}{0.61} &
  0.58 &
  0.58 &
  0.68 &
  0.66 &
  0.63 &
  \cellcolor[HTML]{FFCCC9}{\color[HTML]{000000} \textbf{0.46}} &
  0.57 &
  0.73 &
  0.69 &
  \cellcolor[HTML]{FFCCC9}{\color[HTML]{000000} \textbf{0.47}} \\
 &
  codeparrot \cite{codeparrot} &
  \textit{gpt-2} &
  1.5B &
  \multicolumn{1}{c|}{0.71} &
  0.67 &
  0.67 &
  \cellcolor[HTML]{CACDFA}\textbf{0.80} &
  0.76 &
  0.70 &
  0.59 &
  0.66 &
  \cellcolor[HTML]{CACDFA}\textbf{0.82} &
  0.74 &
  0.64 \\
 &
  codegen-350M-mono\cite{Nijkamp2022ACP}&
  \textit{codegen-350M-multi} &
  350M &
  \multicolumn{1}{c|}{0.73} &
  0.68 &
  0.76 &
  0.78 &
  0.76 &
  0.73 &
  0.57 &
  0.68 &
  \cellcolor[HTML]{CACDFA}\textbf{0.86} &
  0.77 &
  0.58 \\
\multirow{-4}{*}{\textbf{\begin{tabular}[c]{@{}l@{}}Mono-\\ Language\end{tabular}}} &
  codegen-2B-mono\cite{Nijkamp2022ACP}&
  \textit{codegen-2B-multi} &
  2B &
  \multicolumn{1}{c|}{{\ul \textbf{0.84}}} &
  0.79 &
  \cellcolor[HTML]{CACDFA}\textbf{0.84} &
  \cellcolor[HTML]{CACDFA}\textbf{0.90} &
  \cellcolor[HTML]{CACDFA}\textbf{0.85} &
  \cellcolor[HTML]{CACDFA}\textbf{0.81} &
  0.73 &
  \cellcolor[HTML]{CACDFA}\textbf{0.82} &
  \cellcolor[HTML]{CACDFA}\textbf{0.94} &
  \cellcolor[HTML]{CACDFA}{\color[HTML]{000000} \textbf{0.85}} &
  \cellcolor[HTML]{CACDFA}\textbf{0.83} \\ \bottomrule
\end{tabular}

}

\vspace{-0.45cm}

\end{table*} 

\subsection{\ref{rq:asteval} The \asceval Empirical Methodology} 
To answer \ref{rq:asteval}, we generated the normalized log-probabilities (see Sec.\ref{sec:background}) or Next Token Predictions (\ntp) for each code snippet in $\mathcal{S}=$\galeras. These log-probabilities were extracted at inference time for each token position for the 12 \llms. The log-probabilities distributions have a vector size of $|V|$ for each token position in $s \in \mathcal{S}$. These distributions are processed to obtain the log-probability that actually matches the expected token in a position $i$. Therefore, each token position has an associated prediction value that we save for generating the \ntp sequence. Such Next-token Prediction sequence is the input for the aggregation function $\theta$ that generates the corresponding \asceval values (see Sec.~\ref{sec:approach}). Additionally, we computed the cross-entropy loss of each snippet $s$ in our dataset. To obtain the \asceval \textit{Global} value in Tab.~\ref{tab:models} and Fig.~\ref{fig:asc_performance}, we aggregated \asceval performance values (\ie all available \asc) by \llm. The values per model are bootstrapped with the median (size of 500 samplings) to enable a fair comparison among models. Similarly, to obtain the \asceval per Abstract Syntax Concept Category (\eg Data Str, Decision, or Scope), we globally aggregated performance values of tokens under these categories. We also explored with Type Model aggregations (see Table.~\ref{tab:models}).   

\subsection{\ref{rq:astxplainer} The \asccausal Empirical Methodology} 

To answer \ref{rq:astxplainer}, we compute both Pearson correlation $\rho$ values and causal treatment effects (see Def~.\ref{def:effect}) for a subset of 14 syntactic concepts (see Fig.~\ref{fig:asc_causal}). Specifically, we propose the treatments ($T$) \textit{Scope}, \textit{Exceptions}, \textit{Operator}, \textit{Decision}, \textit{Data Structures}, \textit{Functional Programming}, \textit{Natural Language}, \textit{Iterative}, \textit{Types}, and \textit{Testing}. Each \asc was correlated to 4 confounding variables (\ie Cyclo, AST Levels, \#AST Nodes, and Sequence Size) and the cross-entropy loss of \gptI and \monoIIII. We decided to explore only edge case \llms (\ie the best and worst models by \asceval performance) since we detected that the correlated values were very similar across \llms. On the other hand, we estimated the probability of the treatment effect $p(Y|do(T)$ for each \asc and the cross-entropy loss by controlling the 4 previously mentioned confounders. This probability function was estimated using the \textit{doWhy} tool \cite{Sharma2021DoWhyAssumptions}. Table \ref{tab:correlations} summarizes the treatment effects and correlations between the \asceval values locally aggregated (see Sec.~\ref{sec:approach}) and the cross-entropy loss grouped by concept categories ($\mathcal{H}$). 

\subsection{\ref{rq:usability} Qualitative User-Study Methodology}

To answer \ref{rq:usability}, we designed four surveys to understand the perception of software practitioners in regard to the \textit{usability} of \asceval and \ascviz. Our goal is to assess the effectiveness of our \asceval and \ascviz approaches to explain why and how certain source code tokens are predicted by \llms trained on code. Leveraging interpretability techniques to explain the decisions of such models can give software practitioners insights into the behavior and the quality of the predictions.

We introduced a set of code exemplars with their corresponding \ascviz explanation. We asked the participants to rate the explanations for four Python samples distributed across treatments. We use a within-subjects design or repeated measures design, in which every individual receives each of the experimental treatments consecutively. Table \ref{tab:survey} contains a summary with the description of each survey. Each individual survey has two sections. The first section of each survey intends to gauge the proficiency of participants using Python and their familiarity with language models for code generation tasks. Participants were also asked about their knowledge of representation of algorithms (AST) and the major problems that they have faced when using \llms for source code generation. 

In the second section, we provide four Python prompts with an incomplete method along with the prediction of the missing lines given by an \llm. Since our goal is to evaluate the usability of \astxplainer rather than the perception of the participant in regards to the  model performance, we omit details about the model used for predictions (\gptII). Each prompt is accompanied by a visualization (\ie AST-partial, AST-complete, and sequence) that shows the \asceval and \ntp values for the predicted tokens. Then we ask the participant to assess the visualization and rate its usefulness. The visualizations are separated into different surveys. 

Figure \ref{fig:local_evaluation} poses an example of a local evaluation of a code completion task that was presented to the participants. The survey asks to evaluate four different samples processed with \asceval. Some of the samples have syntactic errors. Each survey comprises an incomplete Python method (\ie prompt) and a complete method with the highlighted portion of the generated code (region \circled{1} in Fig.~\ref{fig:local_evaluation}). For each sample in the surveys, we presented a specific visualization. For instance, surveys ($S2$) and ($S3$) contain an AST-based representation similar to the one in Fig.~\ref{fig:local_evaluation}. The AST-complete visualization $S3$ shows the terminal and non-terminal nodes (region \circled{2} in Fig.~\ref{fig:local_evaluation}). Nonetheless, the AST-partial visualization ($S2$) only shows the non-terminal nodes. Finally, a sequence-based visualization of the generated logits for each token was presented to the participants in the survey ($S1$) (region \circled{3} in Fig.~\ref{fig:local_evaluation}).

\section{Results \& Discussion}
\label{sec:results}

\subsection{\ref{rq:asteval} Empirical \asc Performance Evaluation}

In this RQ, we provide an empirical value  (bootstrapped median columns in  Tab.~\ref{tab:models}) of the prediction of Abstract Syntax Concepts for the 12 \llms. We set a threshold of $0.6$ as an acceptable rate of prediction confidence for our \asceval metric. Fig.~\ref{fig:largeTreeMap}, for example, shows our best and worst \llms, \monoIIII and \gptI respectively, at every proposed Abstract Syntax Concept. We observe that, in general, scaling the parameters of \llms plays a fundamental role in the prediction of \asc. The dashed green boxes show the largest \asceval performance increments from the worst to the best concepts. Particularly, \textit{Exceptions}, \textit{Natural Language}, \textit{Operators}, \textit{Types}, and \textit{Decisions} present the biggest jumps in syntactic \asceval performance. 

Our empirical evaluation shows that \asc categories that fulfill the $0.6$ threshold for the 12 \llms are \textit{Scope} with the highest \asceval performance of $0.94$ for the \monolang models, \textit{Iterations} with $0.82$ for \codegenII, and \textit{Testing} with $0.85$ for \monoIIII (see ~Table.\ref{tab:models}). Conversely, we found some concept categories struggle with \asceval performance. We refer to these categories as \textit{erroneous} since they are below $0.5$. Those categories are mainly \textit{Natural Language} category with the largest average median of $0.46$ and \textit{Data Types} with the largest average median of $0.47$ for \nlgpt.

\begin{figure}[ht]
		\centering
  \vspace{-1em}
		\includegraphics[width = 0.5\textwidth]{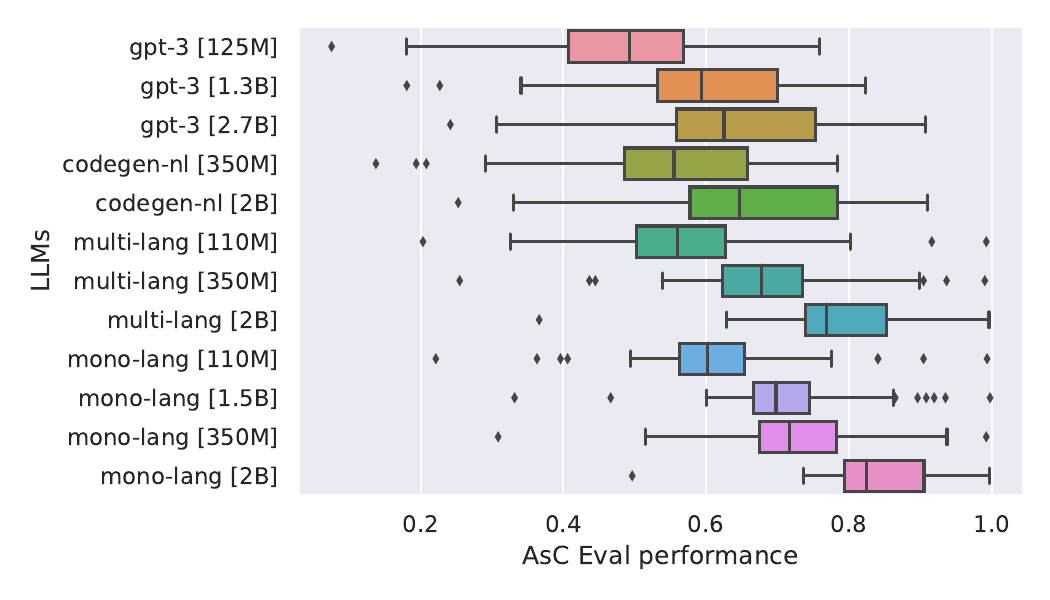}
  \centering
		\caption{\asceval Performance grouped by specific \llms and \asceval density by Model Type.}
        \label{fig:asc_performance}
        \vspace{-1em}
\end{figure}


We believe that models poorly behave with low \asceval performance because category concepts such as \textit{Natural Language} and \textit{Data Types} require more context to be accurately predicted. For instance, the \texttt{\small `string'} concept requires a larger window context before properly being predicted. Similarly, the category \textit{Data Types} is prone to be erroneous since they may appear more frequently at the beginning of the snippets compared to other categories. Also, bear in mind that \textit{Data Types} are less frequent concepts due to the dynamic typing for Python. In general, none of the evaluated architectures performed well at predicting \textit{Data Types} accurately except by \monoIIII, which was trained with a large number of code samples.

Table~\ref{tab:models} depicts that \textit{Iteration} category mostly surpasses the threshold for all our models except for \textit{codeparrot-small-multi} with an average median \asceval of $0.6$. From our smaller models (\ie in a range of millions of parameters), the lowest average median obtained for \gptI is $0.74$, which also surpasses the threshold. This outstanding behavior of \nlgpt models could be explained as Python reserved words for iterations such as \textbf{for} and \textbf{while} also appear in natural language with similar semantics.

Fig.~\ref{fig:asc_performance} indicates that models trained on natural language have more median variability than models fine-tuned on code datasets. For instance, \nlgpt and \nlcodegen report values in a range from $0.2$ to $0.9$. Conversely, fine-tuned models with code such as \monolang has a lower variability than \nlgpt and \nlcodegen categories. For example, \monoIIII has a global avg. median \asceval of $0.84$ and a variability range between $0.7$ and $1.0$, outperforming the $0.6$ threshold. Furthermore,  \monoIIII is our best model with an average global \asceval of $0.84$. On one hand, this suggests that fine-tuned models on code are predicting \asc with higher confidence than natural language-only models. On the other hand, although \multilang models exhibit high variability (from  $0.5$ to $0.9$), their average median \asceval (\ie $0.68$ for \multiI) is even better than natural language models (\ie $0.48$ with variability from $0.2$ to $0.8$ for \gptI).

\begin{boxK}
\textit{\ref{rq:asteval} \asceval:} 
The prediction of syntactic structures highly depends on LLMs' parameter size and fine-tuning strategy. More specifically our largest evaluated model \monoIIII, which was fine-tuned with the BigPython and BigQuery datasets, obtains the highest global average \asc Performace of $0.84$ with the lowest variability.
\end{boxK}

\subsection{\ref{rq:astxplainer} Empirical Causal Evaluation}
In this research question, we want to quantitatively demonstrate that cross-entropy loss of \llms tends to be negatively impacted by the \asceval values at snippet granularity. Therefore, we can explain in lower granularity which parts of the code \llm perform poorly (see red boxes in Tab.~\ref{tab:models}). We showcase empirical evidence that the previous statement holds for both correlation $\rho$ and causal effect $p(y|do(t))$ values. For example, Table~\ref{tab:correlations} shows that, in general, all Abstract Syntax Concept Categories (\ie Global Avg. \asceval) influence the cross-entropy loss for our best (\ie \monoIIII) and worst (\ie \gptI) models, with an average treatment effect of $1.78$ and $-1.60$ respectively. 

The most outstanding finding is that the \textit{Natural Language} category has the largest impact on the cross-entropy loss. For example, the \asc concept \texttt{\small `identifier'} has a causal effect of $-1.78$ for \gptI and $-2.89$ for \monoIIII. In contrast, \textit{Functional Programming} categories present the lowest impact on cross-entropy loss with a subtle \texttt{\small `lambda'} positive causal effect of $0.2$ for \gptI. This subtle positive effect was expected as NL-based \llms have not been fine-tuned on code corpora with \texttt{\small `lambda'} expressions. In addition, we want to highlight the moderate Pearson correlation value between the \texttt{\small `if statement'} concept and the cyclomatic complexity for our best and worst models with the same value of $\rho=0.58$. This observation is consistent with the definition of the cyclomatic complexity metric since this metric takes into consideration the control flows induced by conditional structures.

\begin{table*}[ht]
\centering
\caption{\asccausal: Correlations and Causal Effects.}
\vspace{-0.1em}
\label{tab:correlations}

\scalebox{0.85}{

\setlength{\tabcolsep}{4pt} 

\begin{tabular}{@{}ll|rrrrrrrr|rrrr@{}}
\toprule
\multicolumn{2}{c|}{\textbf{Variable}} &
  \multicolumn{2}{c}{\textbf{Cyclo Complexity}} &
  \multicolumn{2}{c}{\textbf{AST Levels}} &
  \multicolumn{2}{c}{\textbf{\# AST Nodes}} &
  \multicolumn{2}{c|}{\textbf{Sequence Size}} &
  \multicolumn{4}{c}{\textbf{Cross-Entropy Loss}} \\ \midrule
\multicolumn{2}{c|}{\textbf{LLMs}} &
  \multicolumn{1}{c}{\textbf{gpt-125}} &
  \multicolumn{1}{c}{\textbf{mono-2B}} &
  \multicolumn{1}{c}{\textbf{gpt-125}} &
  \multicolumn{1}{c}{\textbf{mono-2B}} &
  \multicolumn{1}{c}{\textbf{gpt-125}} &
  \multicolumn{1}{c}{\textbf{mono-2B}} &
  \multicolumn{1}{c}{\textbf{gpt-125}} &
  \multicolumn{1}{c|}{\textbf{mono-2B}} &
  \multicolumn{2}{c}{\textbf{gpt-125}} &
  \multicolumn{2}{c}{\textbf{mono-2B}} \\ \midrule
\multicolumn{1}{c}{\textbf{Category $\mathcal{H}$}} &
  \multicolumn{1}{c|}{\textbf{AsC}} &
  \multicolumn{8}{c|}{\textbf{$\rho = p(Y|T)$}} &
  \multicolumn{1}{c}{\textbf{$\rho$}} &
  \multicolumn{1}{c}{\textbf{Causal Eff.}} &
  \multicolumn{1}{c}{\textbf{$\rho$}} &
  \multicolumn{1}{c}{\textbf{Causal Eff.}} \\ \midrule
\textit{\textbf{Global}} &
  \textit{Avg. \asceval} &
  \cellcolor[HTML]{EFEFEF}\textbf{0.61} &
  \cellcolor[HTML]{EFEFEF}\textbf{0.60} &
  \cellcolor[HTML]{EFEFEF}\textbf{0.41} &
  \cellcolor[HTML]{EFEFEF}\textbf{0.45} &
  \cellcolor[HTML]{EFEFEF}\textbf{0.51} &
  \cellcolor[HTML]{EFEFEF}\textbf{0.47} &
  \cellcolor[HTML]{EFEFEF}\textbf{0.53} &
  \cellcolor[HTML]{EFEFEF}\textbf{0.48} &
  -0.38 &
  \cellcolor[HTML]{EFEFEF}\textbf{-1.60} &
  -0.38 &
  \cellcolor[HTML]{EFEFEF}\textbf{-1.78} \\ \midrule
 &
  \textit{for\_statement} &
  0.32 &
  0.32 &
  0.30 &
  0.30 &
  0.18 &
  0.19 &
  0.23 &
  0.19 &
  -0.16 &
  -0.10 &
  -0.07 &
  -0.01 \\
\multirow{-2}{*}{\textit{\textbf{Iterative}}} &
  \textit{while\_statement} &
  0.13 &
  0.13 &
  0.08 &
  0.07 &
  0.05 &
  0.05 &
  0.07 &
  0.05 &
  -0.05 &
  -0.11 &
  -0.03 &
  -0.08 \\ \midrule
 &
  \textit{identifier} &
  0.10 &
  0.12 &
  0.10 &
  0.20 &
  \cellcolor[HTML]{EFEFEF}\textbf{0.51} &
  \cellcolor[HTML]{EFEFEF}\textbf{0.46} &
  \cellcolor[HTML]{EFEFEF}\textbf{0.49} &
  \cellcolor[HTML]{EFEFEF}\textbf{0.49} &
  \cellcolor[HTML]{EFEFEF}\textbf{-0.56} &
  \cellcolor[HTML]{EFEFEF}\textbf{-1.78} &
  \cellcolor[HTML]{EFEFEF}\textbf{-0.80} &
  \cellcolor[HTML]{EFEFEF}\textbf{-2.89} \\
\multirow{-2}{*}{\textit{\textbf{\begin{tabular}[c]{@{}l@{}}Natural \\ Language\end{tabular}}}} &
  \textit{string} &
  0.02 &
  0.04 &
  0.20 &
  0.24 &
  0.36 &
  0.30 &
  0.36 &
  0.30 &
  -0.31 &
  -0.36 &
  \cellcolor[HTML]{EFEFEF}\textbf{-0.43} &
  \cellcolor[HTML]{EFEFEF}\textbf{-0.55} \\ \midrule
 &
  \textit{{]}} &
  0.16 &
  0.14 &
  0.21 &
  0.28 &
  0.35 &
  0.36 &
  0.29 &
  0.32 &
  -0.16 &
  -0.04 &
  -0.22 &
  -0.10 \\
\multirow{-3}{*}{\textit{\textbf{Scope}}} &
  \textit{)} &
  0.09 &
  0.03 &
  0.07 &
  0.12 &
  0.37 &
  0.23 &
  0.31 &
  0.23 &
  -0.37 &
  -0.85 &
  \cellcolor[HTML]{EFEFEF}\textbf{-0.54} &
  \cellcolor[HTML]{EFEFEF}\textbf{-1.49} \\ \midrule
\textit{\textbf{Decision}} &
  \textit{if\_statement} &
  \cellcolor[HTML]{EFEFEF}\textbf{0.58} &
  \cellcolor[HTML]{EFEFEF}\textbf{0.58} &
  0.29 &
  0.28 &
  0.20 &
  0.21 &
  0.27 &
  0.23 &
  -0.22 &
  -0.21 &
  -0.11 &
  -0.11 \\ \midrule
 &
  \textit{comparison\_operator} &
  0.34 &
  0.38 &
  0.18 &
  0.20 &
  0.30 &
  0.30 &
  0.30 &
  0.32 &
  -0.13 &
  0.02 &
  -0.11 &
  0.00 \\
\multirow{-2}{*}{\textit{\textbf{Operator}}} &
  \textit{boolean\_operator} &
  \cellcolor[HTML]{EFEFEF}\textbf{0.53} &
  \cellcolor[HTML]{EFEFEF}\textbf{0.54} &
  0.20 &
  0.21 &
  0.17 &
  0.18 &
  0.22 &
  0.20 &
  -0.10 &
  0.01 &
  -0.08 &
  -0.09 \\ \midrule
 &
  \textit{for\_in\_clause} &
  0.28 &
  0.29 &
  0.18 &
  0.18 &
  0.15 &
  0.14 &
  0.13 &
  0.13 &
  -0.03 &
  0.09 &
  -0.03 &
  0.04 \\
 &
  \textit{if\_clause} &
  0.20 &
  0.22 &
  0.10 &
  0.10 &
  0.06 &
  0.06 &
  0.07 &
  0.06 &
  -0.01 &
  0.19 &
  0.01 &
  0.13 \\
 \multirow{-3}{*}{\textit{\textbf{\begin{tabular}[c]{@{}l@{}}Functional \\ Programming\end{tabular}}}} &
  \textit{lambda} &
  -0.01 &
  -0.01 &
  0.12 &
  0.12 &
  0.18 &
  0.18 &
  0.16 &
  0.16 &
  -0.04 &
  0.20 &
  -0.05 &
  0.06 \\
   \bottomrule
\end{tabular}

} 
\end{table*}

 

\begin{boxK}
\textit{\ref{rq:astxplainer} \asccausal: }
We can observe that cross-entropy loss of LLMs tends to be negatively impacted by the \asceval values at snippet granularity.
For instance, 
\textit{identifiers} affect both \gptIII and \monoIIII cross-entropy loss with Average Treatment Effects of $-1.78$ and $-2.89$ respectively.
\end{boxK}

\subsection{\ref{rq:usability} User-Study on AsC Visualization Results}
In this RQ, we evaluate our \asceval and \ascviz in a practical scenario. We assessed how useful is the AST visualization for practitioners after collecting a total of $40$ responses (with sampling confidence of $90\%$ and a margin error of $15\%$). Each participant was randomly and uniformly assigned to either the control or one of the survey groups. 

\begin{table}[h]
\vspace{-0.1cm}
\begin{center}
\caption{User study visualization results}
\scalebox{0.95}{%

\setlength{\tabcolsep}{3pt} 

\begin{tabular}{lll|lccc}
\hline
\textbf{ID} &
  \textbf{Group} &
  \textbf{Visualization} &
  \multicolumn{4}{c}{\textit{\textbf{User study visualization results}}} \\ \hline
\textit{C} &
  \textit{Control} &
  \begin{tabular}[c]{@{}l@{}}Prompt\\ predicted tokens.\end{tabular} &
  \textbf{Agreement} &
  \textbf{\begin{tabular}[c]{@{}c@{}}\%\\ useful\end{tabular}} &
  \textbf{\begin{tabular}[c]{@{}c@{}}\% \\ readable\end{tabular}} &
  \textbf{\begin{tabular}[c]{@{}c@{}}\% \\ complex\end{tabular}} \\ \hline
\multirow{4}{*}{$S_1$} &
  \multirow{4}{*}{\textit{Sequence}} &
  \multirow{4}{*}{\begin{tabular}[c]{@{}l@{}}Prompt\\ predicted tokens\\ NTP.\end{tabular}} &
  Strongly &
  14 &
  0 &
  0 \\
 &
   &
   &
  Agree &
  \textbf{43} &
  29 &
  \textbf{43} \\
 &
   &
   &
  Neutral &
  29 &
  29 &
  29 \\
 &
   &
   &
  Disagree &
  14 &
  \textbf{43} &
  29 \\ \hline
\multirow{4}{*}{$S_2$} &
  \multirow{4}{*}{\textit{\begin{tabular}[c]{@{}l@{}}AST\\ partial\end{tabular}}} &
  \multirow{4}{*}{\begin{tabular}[c]{@{}l@{}}Prompt\\ predicted tokens\\ partial AST\\ NTP.\end{tabular}} &
  Strongly &
  14 &
  14 &
  0 \\
 &
   &
   &
  Agree &
  \textbf{29} &
  \textbf{43} &
  14 \\
 &
   &
   &
  Neutral &
  \textbf{29} &
  29 &
  \textbf{43} \\
 &
   &
   &
  Disagree &
  \textbf{29} &
  14* &
  \textbf{43} \\ \hline
\multirow{4}{*}{$S_3$} &
  \multirow{4}{*}{\textit{\begin{tabular}[c]{@{}l@{}}AST\\ complete\end{tabular}}} &
  \multirow{4}{*}{\begin{tabular}[c]{@{}l@{}}Prompt\\ predicted tokens\\ complete AST\\ NTP.\end{tabular}} &
  Strongly &
  0 &
  0 &
  17 \\
 &
   &
   &
  Agree &
  \textbf{50} &
  17 &
  \textbf{50} \\
 &
   &
   &
  Neutral &
  33 &
  33 &
  17 \\
 &
   &
   &
  Disagree &
  17 &
  \textbf{50} &
  17 \\ \hline
\end{tabular}
}%

\label{tab:survey}
\end{center}
\small{The * depicts the only \textit{``Strongly Disagree''} reported in the survey}
\end{table} 

Our user study (see Tab.~\ref{tab:survey}) revealed that a partial AST visualization (\ie AST with only non-terminal nodes) is preferred over the complete AST representation ($S_3$) and sequence visualization ($S_1$). Furthermore, AST partial visualization ($S_2$) is found to be particularly less complex by participants with a proportion of $43\%$ in the survey compared to the control $C$. However, AST complete visualization ($S_3$) is perceived to be more complex by participants with a proportion of $67\%$ compared to the control. However, we observed that $43\%$ of participants would like to use the sequential visualization rather than not having any visual help (\ie control group $C$) and only $29\%$ of participants considered sequential visualization easy to read and use. Surprisingly, $57\%$ of participants found the sequence visualization to be highly useful in explaining the behavior of the model. This preference can be explained by considering the fact that only 42\% of the participants have formal education in ML. 

In all the surveys, participants agreed that visualizing \textit{Error} nodes in \ascviz with their corresponding \asceval value was very useful to understand the long-range dependencies of the generative process. The participants highlight the fact that \ascviz is useful to explain why the \llm generates an erroneous prediction. More specifically, the participants mentioned that the color code for the \asceval values helps them to differentiate between predicted probabilities. The participants also indicated that they would like the visualizations to include information about the workflow of the program.

\begin{boxK}
\textit{\ref{rq:usability} Usability of \asceval:}
We found that the partial visualization of the AST is the most \textit{readable} representation to showcase local aggregated predictions with 57\% agreement. Although AST partial visualization has mixed opinions about its \textit{usefulness} with an agreement rate of 29\%, the AST complete visualization has an agreement rate of 50\%.
\end{boxK}

\section{Conclusion \& Future Work}
\label{sec:conclusion}

Our research proposes an Abstract Syntax Concept approach to evaluate and explain the performance of Large Language Models for code. We conducted a rigorous empirical evaluation on 12 popular \llms using a curated dataset that uncovered novel performance qualities of \llms for code. Our empirical evaluation revealed that mono-language outperforms multi-language \llms at predicting all types of syntax concepts. This suggests the importance of fine-tuning strategies over model size. In addition, we demonstrated that Abstract Syntax Concepts influence the cross-entropy loss of \llms after controlling for code confounders. In fact, this influence persists across models at different parameter sizes and fine-tuning strategies. Additionally, we illustrated the utility of visualizations built upon our defined Abstract Syntax Concepts in a user study. We believe these results illustrate the promise of blending explainability and evaluation techniques for \llms of code, and signal the potential for future work to further integrate explainability techniques into future \llm benchmarks.

\bibliographystyle{IEEEtran}
\bibliography{IEEEabrv,references}
\end{document}